\documentclass[notitlepage]{revtex4-1}

\usepackage{graphicx}% Include figure files
\usepackage{dcolumn}% Align table columns on decimal point
\usepackage{bm}% bold math
\usepackage{booktabs}
\usepackage{float}
\usepackage[english]{babel}
\begin{document}

\preprint{APS/123-QED}

\title{Rapid Thermal Annealing for Surface Optimisation of ZnO Substrates for MBE-Grown Oxide Two-Dimensional Electron Gases}% Force line breaks with \\

\author{M.~J.~Sparks}
 \email{matthew.sparks.15@ucl.ac.uk}
 \altaffiliation{London Centre for Nanotechnology, University College London, 17-19 Gordon Street, London WC1H 0AH, United Kingdom}
\author{O.~W.~Kennedy}%
 \altaffiliation{London Centre for Nanotechnology, University College London, 17-19 Gordon Street, London WC1H 0AH, United Kingdom}
\author{P.~A.~Warburton}%
 \altaffiliation{London Centre for Nanotechnology, University College London, 17-19 Gordon Street, London WC1H 0AH, United Kingdom}
 \altaffiliation[Also at ]{Department of Electronic and Electrical Engineering, University College London, Torrington Place, London WC1E 7JE, United Kingdom.}%Lines break automatically or can be forced

\date{\today}% It is always \today, today,

\begin{abstract}
Two-dimensional electron gases (2DEGs) at the ZnO/ZnMgO interface are promising for applications in spintronics and quantum computing due to the combination of low spin-orbit coupling and high electron mobility. Growing high mobility 2DEGs requires high quality substrates with low impurity densities. In this work we demonstrate a ZnO substrate sample treatment combining high temperature rapid thermal annealing and chemical etching to improve the surface quality for the subsequent growth of 2DEGs. This process enables the growth of a 2DEG with low-temperature mobility of $4.8\times10^4$~cm$^2$V$^{-1}$s$^{-1}$. An unannealed control sample shows a scattering rate at least three times greater than the annealed sample.
\end{abstract}

\maketitle

%\tableofcontents

\section{Introduction}

Two-dimensional electron gases (2DEGs) at the ZnO/ZnMgO interface combine exceptional electron mobilities and consequent long diffusion lengths \cite{falson_mgzno/zno_2016} in a low spin-orbit-coupling material \cite{fu_spin-orbit_2008}. This allows, in principle, electrons to be transported long distances whilst maintaining spin coherence \cite{han_spin-polarized_2012}. Furthermore, fractional quantum Hall states in these 2DEGs realise non-abelian statistics \cite{tsukazaki_observation_2010, moore_nonabelions_1991} and could be used to build topological quantum computers \cite{kitaev_fault-tolerant_2003}. These properties mean that this material system has significant promise for spintronics and quantum-computation applications.

High mobility 2DEGs require a low density of scattering centres. These can be formed by crystalline defects \cite{kumar_electron_2020}, ionized donors \cite{umansky_extremely_1997}, alloy scattering \cite{kumar_electron_2020}, electron-electron scattering \cite{lisesivdin_scattering_2007} or atomic impurities \cite{li_electron_2014}.  Minimizing the crystalline defect density requires the use of an epitaxially matched substrate, with the best results found using single-crystal single crystal ZnO \cite{falson_mgzno/zno_2016}. Alloy scattering is small in these systems as the polar interface results in electrons being confined within ZnO \cite{tampo_two-dimensional_2006, ye_two-dimensional_2010}. There is a trade-off to be made between electron density and electron-induced scattering which is determined by the Mg concentration of ZnMgO layers \cite{ye_origin_2013, ye_two-dimensional_2010}. In MBE-grown ZnO/ZnMgO 2DEGs, interface roughness scattering and impurity scattering have been shown to be the dominant factors for electron mobility \cite{kumar_electron_2020}. It is known, for instance, that lithium acts as an acceptor in ZnO and can significantly reduce ZnO conductivity \cite{lin_hydrothermal_2009, monakhov_zinc_2009, svensson_hydrothermally_2011} and harm 2DEG formation \cite{li_electron_2014}. Atomic impurities are known to be present in ZnO substrates and are incorporated during hydrothermal growth \cite{svensson_hydrothermally_2011}. Whilst exceptionally low impurity ZnO crystals have been grown using a platinum-lined autoclave \cite{avrutin_bulk_2010, lin_hydrothermal_2009}, since the decommissioning of this equipment, only ZnO substrates grown in rhenium/iridium-lined autoclaves are available.

Surface band-bending in n-type ZnO means the bands bend upwards at the ZnO surface. This creates an electric field exerting a small electrostatic force toward (away from) the surface on positively (negatively) charged impurities. At room temperature this force is insufficient to cause impurities to move in the sample, however at elevated temperatures impurities can diffuse in ZnO over several microns \cite{monakhov_zinc_2009, svensson_hydrothermally_2011}. The small electric field from band-bending will bias the diffusion of positively charged impurities towards the surface of ZnO. This effect is more pronounced at the negatively charged O-polar face \cite{monakhov_zinc_2009}. A combination of thermal annealing and etching can therefore induce positively charged impurities to diffuse to the surface before being etched away as shown schematically in Figure \textbf{1(a-c)}.

In this work we investigate the combination of substrate annealing and hydrochloric acid etching of hydrothermally grown single crystal ZnO substrates as a preparatory step before 2DEG growth. The annealing stage is designed to cause positively charged impurities to diffuse to the surface, combined with the electric field from surface-band-bending in the intrinsically n-type ZnO \cite{monakhov_zinc_2009}.  We show that the room temperature crystal surface resistance decreases under this treatment. Additionally we use MBE to grow two ZnO/ZnMgO heterostructure bilayers which are identical apart from the fact that one sample has undergone the anneal/etch procedure but the other has not been annealed. We observe that a 2DEG forms in the heat-treated sample but not in the unannealed sample, confirming that the rapid thermal treatment has enhanced 2DEG formation.

\section{Materials \& Methods}

A single 10 $\times$ 10 $\times$ 0.5~mm Zn-face (001) epipolished ZnO crystal sourced from SurfaceNet GmbH was cut into five equally sized pieces. Each piece was degreased in acetone, isopropyl alcohol and  deionised water. Four of the pieces were then annealed by a series of rapid thermal anneals in a Solaris 100 rapid thermal processor system. Each anneal cycle saw the sample raised to 1000$^\circ$C in nitrogen at atmospheric pressure for 5s, before being cooled to 100~$^\circ$C. After annealing the samples were etched in 7:200 HCl:water solution for 30s and then rinsed in DI Water.

These samples were then contacted using a linear array of 300 $\times$ 300~$\mu$m Ti/Au ohmic contacts with a pitch of 500~$\mu$m. These were fabricated onto the Zn-polar face by a lift-off process. Metallisation was performed by DC magnetron sputtering in an SVS 6000 deposition system. Prior to deposition the surface was cleaned using an in-situ argon ion mill. A device schematic is shown in Figure \textbf{1(d)}.

\begin{table}[h]

\centering
%% \tablesize{} %% You can specify the fontsize here, e.g., \tablesize{\footnotesize}. If commented out \small will be used.
\resizebox{0.8\linewidth}{!}{
\begin{tabular}{cccc}
\toprule
\textbf{Sample}	& \textbf{Zn-Flux ($\times10^{-7}$ Torr)} & \textbf{Mg-Flux ($\times10^{-8}$ Torr)} & \textbf{Mg/Zn Flux Ratio}\\
\midrule
 Treated   & 4.8 & 1.8 & 0.038\\
 Untreated & 4.3 & 1.9 & 0.044\\
\bottomrule
\end{tabular}}
\caption{Incident beam fluxes for the treated and untreated heterostructure samples.}
\end{table}

The ZnO/ZnMgO heterostructures were grown on a pair of ZnO substrates, one of which had been subjected to fifteen rapid thermal anneals, followed by a dilute HCl etch. The other sample was not annealed but was etched in HCl. Hereafter these samples are referred to as the `treated' and `untreated' sample, respectively. Heterostructures were grown in an SVTA oxygen plasma assisted MBE system. An initial 100~nm ZnO buffer layer was deposited onto the Zn-polar face of the substrate at 500~$^\circ$C, followed by a 40~nm ZnMgO layer deposited at 750~$^\circ$C. Buffer layers were employed because ZnO homoepitaxial layers have been shown to greatly increase the quality of subsequent ZnMgO growth on ZnO substrates \cite{yuji_mgxzn1-xo_2008, cho_issues_2005}. Table \textbf{1} shows the Zn and Mg fluxes used during growth. Oxygen plasma was maintained at an RF power of 300~W and flow rate of 3~SCCM (chamber pressure of 10$^{-5}$ Torr) throughout the growth of all layers.

Hall bars consisting of a mesa measuring 10~$\mu$m in width were produced by milling the samples to a depth of 200~nm with an argon plasma in the SVS6000 system. These were contacted by Ti/Au leads deposited using the same process as detailed above. Electrical characterisation of the samples was conducted with an LOT Quantum Design physical property measurement system (PPMS) across a temperature range of 2~K to 300~K in a perpendicular magnetic field up to 14~T. Measurements were conducted in a 4-point measurement setup with a quasi-static current bias.

\section{Results}

\subsection{Substrate Thermal Processing}

 By plotting the resistance as a function of contact separation, the substrate surface resistance per unit length $R_L$ can be extracted, as shown in figure \textbf{1d}. This measurement is contact resistance independent. $R_L$ is used as a proxy for sheet resistance. Despite non-uniform current flow in this geometry preventing the sheet resistance being extracted, $R_L$ still gives a good indication of substrate resistivity. We plot $R_L$ as a function of the number of annealing cycles in figure \textbf{1e}. The resistance initially increases before gradually decreasing and reaching a level lower than that of the untreated substrate. The reduction in resistance after many annealing cycles is promising as it suggests these substrates will result in higher 2DEG mobilities \cite{coke_electron_2017}, but the non-monotonic behaviour warrants explanation.

\begin{figure}[h]
\centering
\includegraphics[width=0.8\linewidth]{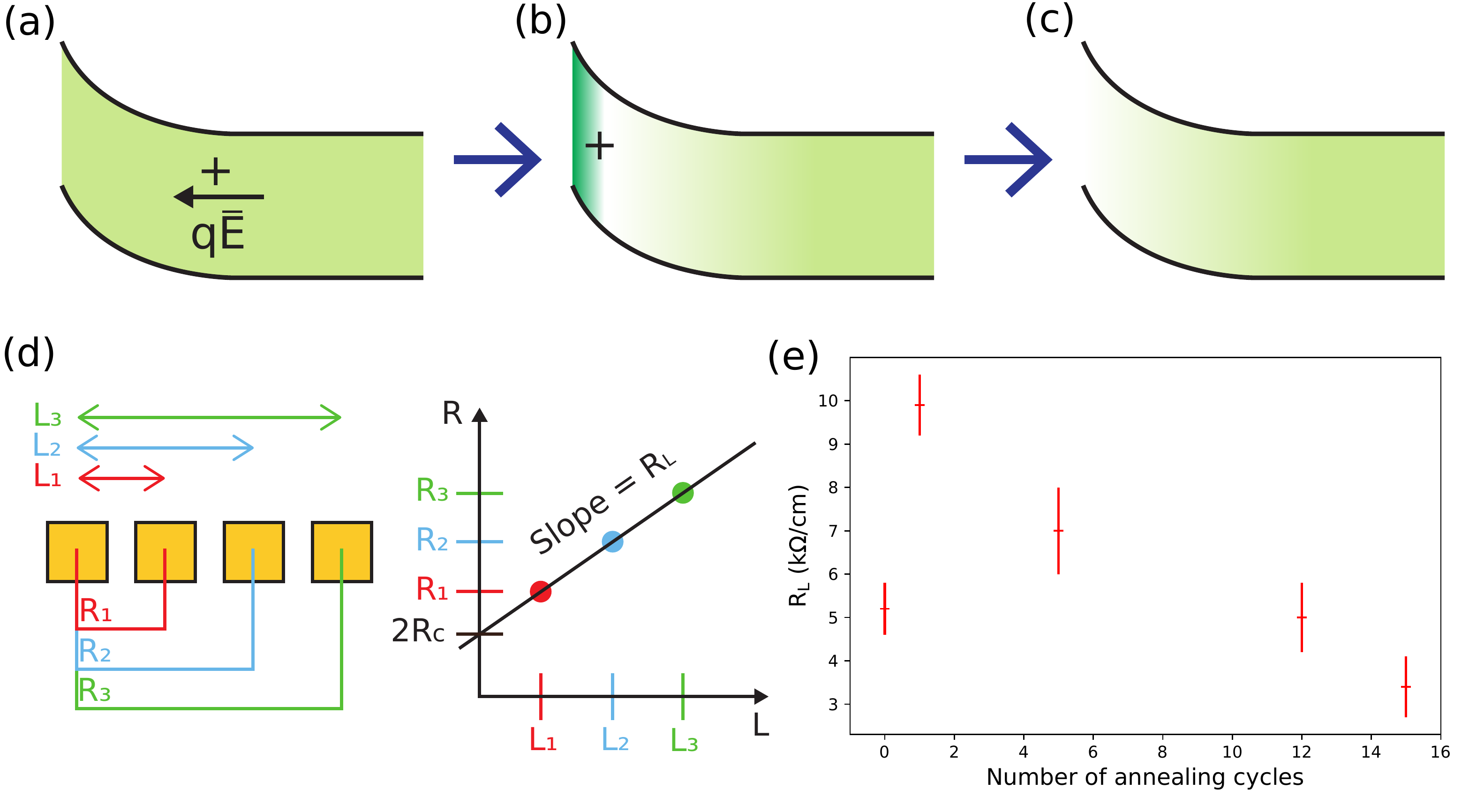}
\caption{Figure a shows the band bending close to the substrate surface. Arrow shows force acting on the impurities with positive charge $q$ in field produced by band bending $\bar{E}$. Uniform impurity distribution is represented by uniform green shaded region between bands. Figure b show impurities congregating at the substrate surface at high temperatures, resulting in a depleted impurity density deeper into the substrate. Figure c shows the system after impurities have been removed, either through surface evaporation or HCl etching. The depleted impurity region is now close to the substrate surface. Figure d shows the measurement setup for the thermally treated substrates as well as a graphical representation of how $R_L$ was measured. Figure (e) shows the ZnO substrate surface resistance per unit length as a function of the number of anneal cycles to a temperature of 1000~$^\circ$C for 5s}
\end{figure}

In n-type ZnO, the initial rise in resistance suggests an increase in acceptor density close to the substrate surface. The drop in resistance with further cycles suggests that these acceptors are removed from the conduction channel. We consider the effects of lithium impurities which have been established as an electrically limiting defect in ZnO \cite{bjorheim_h_2012}. Experiments have shown that heating ZnO with uniformly distributed Li impurities causes the Li to migrate towards the the substrate surface where it congregates \cite{monakhov_zinc_2009} and can then diffuse out of the crystal. Our data suggests that Li density at the surface actually increases as the short anneal cycles are insufficient to allow Li to fully diffuse at the surface. Subsequent anneals allow Li to diffuse to the substrate surface where it may diffuse out of the crystal or be etched away in the HCl etching stage. The diffusivity of Li in ZnO at 1000~$^\circ$C from \cite{knutsen_diffusion_2013} allows us to estimate that Li will diffuse between 6 and 36~$\mu$m per 5~s anneal. This is consistent with surface depletion regions in single-crystal bulk ZnO being a few micrometers \cite{allen_influence_2007}. A single ramp should be sufficient for Li ions to diffuse to the Zn-polar substrate surface, leading to an increase in surface resistance. The Zn-polar face is more resistant to HCl etching than the O-polar face \cite{hupkes_chemical_2012}, so the primary mechanism for Li removal from this face is surface evaporation due to further annealing \cite{svensson_hydrothermally_2011}. This explains the initial rise then fall of R$_L$ as seen in figure \textbf{1d}. Li concentration dynamics could be better established by measuring elemental secondary ion mass spectroscopy (SIMS) both before and after annealing.

\subsection{2DEG Growth and Characterisation}

\begin{figure}[h]
\centering
\includegraphics[width=0.8\linewidth]{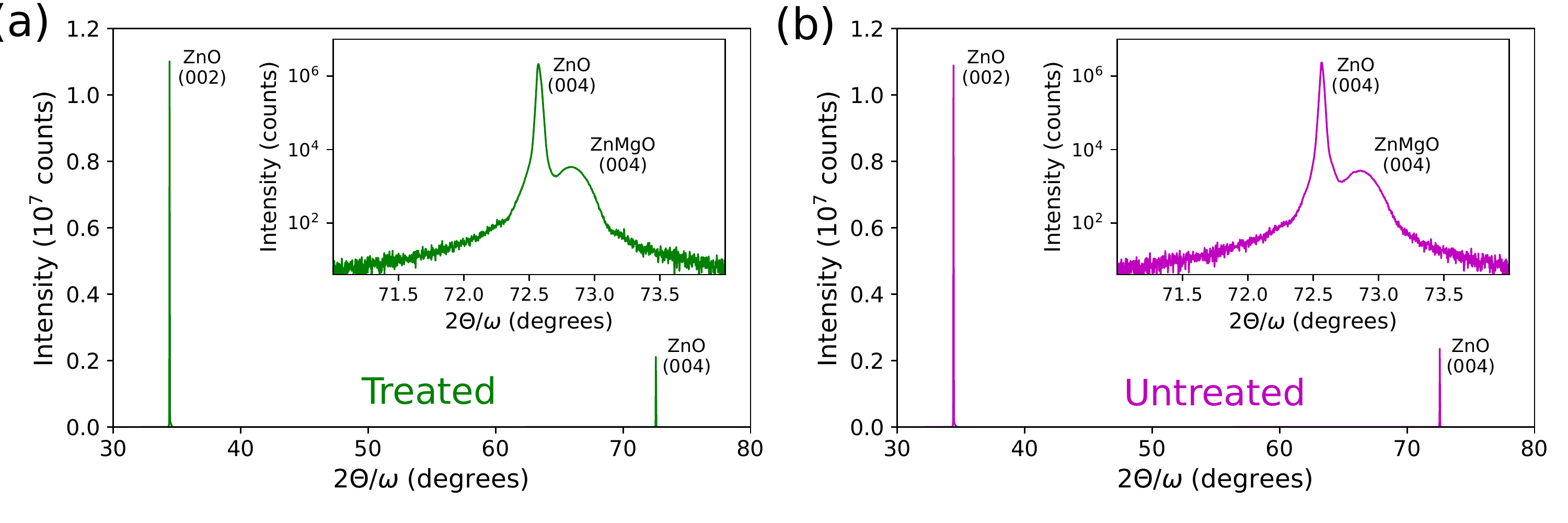}
        \caption{X-ray diffraction plots for ZnO/ZnMgO heterostructures grown on treated (\textbf{a}) and untreated (\textbf{b}) substrates. Insert in XRD plots shows a magnification of the (004) ZnO peak and ZnMgO shoulder peak with a logarithmic intensity scale.}
\end{figure}

Having established a procedure to improve substrate quality by promoting lithium out-diffusion we verify it by using treated substrates for 2DEG growth. Figure \textbf{2a-b} shows the x-ray diffraction (XRD) pattern for the ZnO/ZnMgO heterostructures grown by MBE on treated and untreated substrates. The presence of ZnMgO is confirmed by the shoulders near the ZnO (002) and (004) peaks. Additionally the Mg concentration of the ZnMgO layer can be determined by comparing the location of the ZnO (004) and shoulder peak by using the method outlined in \cite{kozuka_precise_2012}. The Mg content for the treated and untreated sample are 6.1~\% and 7.4~\% respectively.

\begin{figure}[h]
\centering
\includegraphics[width=0.8\linewidth]{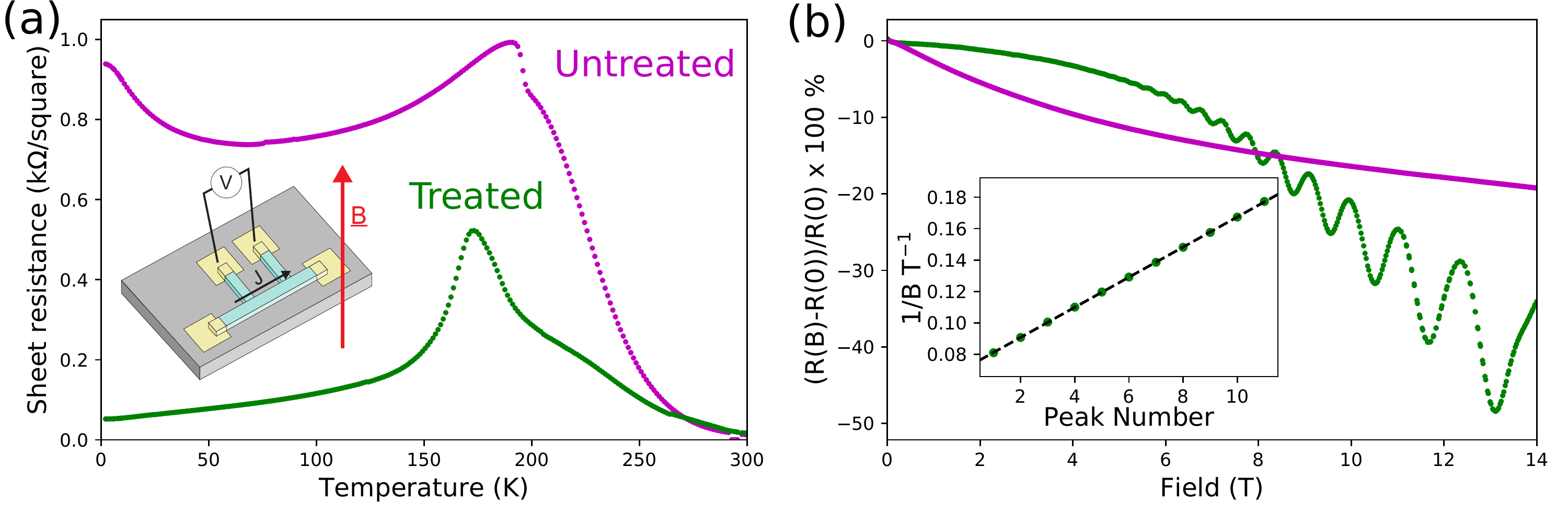}
        \caption{4-point resistance measurement of treated and untreated samples. Figure \textbf{a} shows the zero-field longitudinal sheet resistance vs temperature for ZnO/ZnMgO heterostructures grown on treated (green) and untreated (magenta) substrates. Figure \textbf{b} shows the longitudinal magnetoresistance measurements for treated (green) and untreated (magenta) sample. Insert shows 1/B vs peak number for the treated sample. Magnetic field is taken at the apex of each peak. Least-square linear fit used to determine carrier concentration is shown as dashed line. Measurements were conducted using a 4-point setup at a temperature of 2~K. Magnetic field was applied perpendicular to the substrate surface as shown in the inset in \textbf{a}.}
\end{figure}

Figure \textbf{3a} shows the longitudinal resistance of these samples as a function of temperature as they were cooled from 300~K to 2~K. Both samples show an initial rise in resistance as the temperature falls from 300~K to 200~K, which is typical of semiconductor carrier freeze-out. The resistance reaches a maximum at approximately 160~K before starting to fall. This decrease is attributed to 2DEG formation, as charge carriers are confined to the conduction band quantum-well at the ZnO/ZnMgO interface. Below 100~K the resistance in both samples falls slowly as the 2DEG becomes the dominating conduction path. The resistance of the untreated sample is higher at these temperatures implying more electron scattering in this sample. Below 50~K the behaviour of the samples diverges: the resistance of the treated sample continues to fall but the untreated sample resistance starts to increase with falling temperature. 
%%%This line needs moving later in the paper%%%
%Previous 2DEG studies have demonstrated an increase in resistivity attributed to the electron interference effect \cite{das_effects_2015}. 

Figure \textbf{3b} shows the 2~K longitudinal resistance of the two samples as a function of magnetic field applied normally to the substrate surface. The treated sample (green) shows clear Shubnikov de Haas (SdH) oscillations at fields above 5~T, while no such oscillations are observed in the untreated sample (magenta). The SdH oscillations were used to determine the mobility and carrier concentrations of the 2DEG following \cite{coke_electron_2017}. The insert in \textbf{3d} shows the fit used to determine these properties. The extracted sheet carrier concentration of the treated sample is 5.1 $\times$ 10$^{12}$~cm$^{-2}$ and mobility is 4.8 $\times$ 10$^4$~cm$^2$/Vs at 2~K. This value is three orders of magnitude higher than the low-temperature mobility of ZnO \cite{wagner_halleffekt_1974}.

The untreated sample shows no SDH oscillations but does show a negative magnetoresistance across the entire range of applied magnetic field. This behaviour is similar to that shown by Das et al. \cite{das_effects_2015} for ZnO 2DEGs on c-plane sapphire substrates and is attributed to weak localisation. This suggests that carriers are at least partially confined in the untreated sample; however, unlike the treated sample, defect scattering prevents the observation of SdH oscillations.

For high contrast SdH oscillations, the separation of the Landau levels $\hbar \omega_c$ must exceed the thermal energy $kT$. Defects (including ionised impurities) broaden the Landau levels. The magnitude of this broadening is given by:

\begin{equation}
\Gamma = \hbar \sqrt{\frac{2\omega_c}{\pi \tau_f}}
\end{equation}
where $\tau_f$ is the mean time between electron scattering events \cite{ferry_transport_2009}. Inserting the relation for cyclotron frequency $\omega_c=eB/m^*$ (where $m^*$ is the effective mass of the electrons in the 2DEG) gives an upper bound on the scattering rate:

\begin{equation}
\tau_f^{-1} < \frac{\pi eB}{2m^*}
\end{equation}
We take $m^* = 0.3m_e$, which is typical of values seen in the literature \cite{falson_mgzno/zno_2016, tsukazaki_high_2008}. No SdH oscillations are observed in the unannealed sample up to a field of 14T. From equation (\textbf{4}) we extract a lower bound for $\tau_f^{-1}$ of $1.3 \times 10^{13}$~s$^{-1}$ for the unannealed sample. In the treated sample, SdH oscillations are visible at fields of 5T suggesting a upper bound for $\tau_f^{-1}$ in this sample of $4.6 \times 10^{12}$~s$^{-1}$. Sample treatment has led to at least a factor three decrease in the scattering rate, indicating a reduction in the defect density.

\section{Conclusions}
In summary, we have demonstrated how to exploit rapid thermal annealing to improve the surface quality of ZnO substrates. We observe a non-monotonic response in the substrate surface resistance as a function of the number of applied anneal cycles at 1000~$^\circ$C. We propose a model of defect surface-migration and out-diffusion of impurities with successive anneal cycles. We use this substrate treatment to successfully grow high-mobility 2DEGs and show that the untreated control sample shows no 2DEG. The lack of 2DEG in the unannealed sample is attributed to a higher defect density in the sample. Rapid thermal annealing combined with surface etching has led to at least a factor three reduction of the scattering rate in our samples.

\bibliography{MDPI-Paper-arXiv}% Produces the bibliography via BibTeX.

\end{document}